\documentclass[12pt]{article}
\usepackage[margin=1.in]{geometry}
\usepackage[utf8]{inputenc}
\usepackage{amsthm,amsmath,physics,mathtools,amssymb}

\usepackage{charter}

\usepackage{CJKutf8}

\usepackage{enumitem}

\usepackage{hyperref}
\hypersetup{
    colorlinks=true,
    linkcolor=blue,
    filecolor=magenta,      
    urlcolor=cyan,
}
\usepackage[capitalise]{cleveref}

\theoremstyle{definition}

\usepackage{graphicx}
\graphicspath{ {images/} }

\usepackage{tensor}

\DeclarePairedDelimiterX{\inp}[2]{\langle}{\rangle}{#1, #2}

\usepackage{xparse}

\NewDocumentCommand\LH{mo}{%
  \IfNoValueTF{#2}
   {\mathcal{L}(\mathcal{H}^{#1})}
   {\mathcal{L}(\mathcal{H}^{#1},\mathcal{H}^{#2})}%
}

\newcommand\id{\leavevmode\hbox{\small1\kern-3.3pt\normalsize1}}

\allowdisplaybreaks

\usepackage{bbold}

\usepackage{authblk}

\usepackage[title]{appendix}

\usepackage{mciteplus}

\usepackage{enumitem} 

\title{Is singularity resolution trivial?}

\author{Ding Jia (贾丁)\thanks{djia@perimeterinstitute.ca}}
\affil{Perimeter Institute for Theoretical Physics, Waterloo, Ontario, N2L 2Y5, Canada}
\affil{Department of Physics and Astronomy, University of Waterloo, Waterloo, Ontario, N2L 3G1, Canada}
\date{}

\begin{document}

\begin{CJK*}{UTF8}{gbsn}
\maketitle
\end{CJK*}

\begin{abstract}
Many non-trivial ideas have been proposed to resolve singularities in quantum gravity. In this short note I argue that singularity resolution can be trivial in gravitational path integrals, because geodesically incomplete singular spacetimes are usually not included in the sum. For theories where this holds, there is no need to develop non-trivial ideas on singularity resolution. Instead, efforts should better be directed to understand tunneling processes and complex-valued spacetimes.
\end{abstract}

\section{Is singularity resolution non-trivial?}\label{sec:i}

Spacetime singularities mark the breakdown of General Relativity and form a major motivation for quantum gravity. Many ingenious ideas have been considered before to resolve singularities in quantum gravity. In some proposals it is important to make a judicious choice of variable, e.g., using loop variables as opposed to the metric variable \cite{Ashtekar2006QuantumInvestigation, Ashtekar2009SingularityOverview}. In some proposals it is important to choose the right action, e.g., add higher order terms to the Einstein-Hilbert action \cite{Borissova2020TowardsIntegral}. In some proposals it is important to pick the right boundary condition, e.g., impose a special final boundary condition at the singularities \cite{Horowitz2003TheState}. In some proposals it is important to understand singular solutions at a detailed level, e.g., quantize starting from BKL type solutions \cite{PerryFutureParadox}. These and other proposals leave the impression that singularity resolution in quantum gravity is non-trivial and requires some ingenious input.

As a counterpoint this short note presents a trivial idea on singularity resolution. I argue that generically spacetime configurations singular in the sense of geodesic incompleteness simply do not belong to gravitational path integrals. For theories where this holds, singularities are resolved trivially, and ingenious endeavours can be saved. Instead, efforts should be directed to understanding tunneling processes and complex solutions, as I shall explain below.

\section{Singular configurations fall out}\label{sec:scdnb}

\begin{figure}
    \centering
    \includegraphics[width=.7\textwidth]{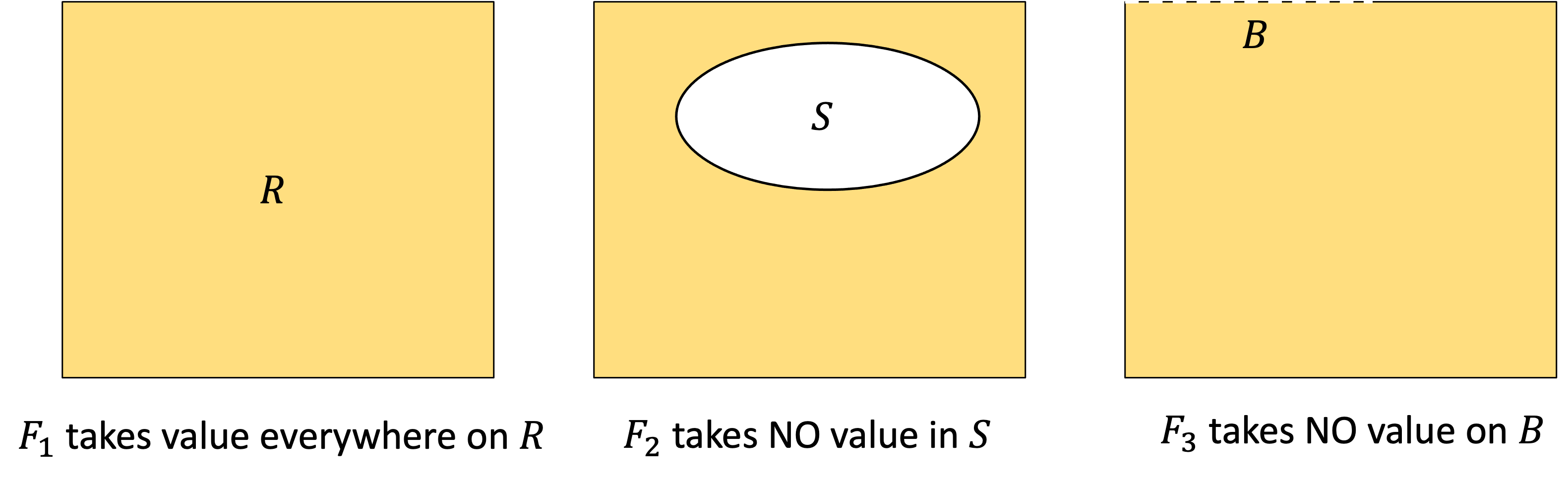}
    \caption{Matter configurations $F_2$ and $F_3$ do not assume values in the interior or on the boundary of the region $R$, so only $F_1$, which assumes value everywhere, belongs to the matter path integral on that region.}
    \label{fig:spic}
\end{figure} 

In a matter path integral on a classical spacetime, we sum only over matter configurations which assumes values everywhere in the region that the path integral refers to (\Cref{fig:spic}). It is natural to expect that the same holds for gravitational path integrals. Yet a geodesically incomplete singular configuration does not meet this requirement. For such a configuration we could follow a causal curve for some finite affine parameter and find out that the gravitational variable stops possessing values. If gravitational path integrals are similar to the matter ones in this respect, singular configurations do not belong to the sum. Consequently the singular spacetimes relevant to the classical theory according to the singularity theorems \cite{wald2010general} become irrelevant to the quantum theories defined by path integrals.

This plausibility argument may not hold for every theory of quantum gravity
, but can be checked to hold in some explicitly defined gravitational path integrals. For example, causal dynamical triangulation \cite{Ambjorn2012NonperturbativeGravity}, locally causal dynamical triangulation \cite{Jordan2013CausalFoliation, Jordan2013DeFoliation} and Lorentzian simplicial quantum gravity \cite{Jia2022ComplexProspects} with a lightcone constraint \cite{JiaLightGravity} are defined by path integrals over piecewise flat spacetime configurations. In all these theories, any point in the interior of a configuration has two lightcones attached to it. A causal path reaching an interior point from one lightcone can always be extended away through the other lightcone. Therefore a causal path either extends indefinitely, or terminates when it reaches the boundary of the simplicial manifold if there is one.\footnote{In the latter case the gravitational variables (edge squared lengths) assume values on the boundary, so the boundary cannot be interpreted as a singularity.} To the extent that singular configurations are characterized by the inextendability of causal paths, they are simply not included in the path integrals.\footnote{In variants of the above theories, a point of a spacetime configuration is allowed to have more or fewer than two lightcones \cite{Loll2006SumGravity, Sorkin1990CONSEQUENCESTOPOLOGY, SorkinLorentzianVectors}. If a point has just one lightcone, a causal path may not be extendable beyond it. This is usually interpreted as due to topology change rather than singularities to be avoided.}

Similar checks can be performed on other explicitly defined gravitational path integrals. While it is possible to encounter peculiar cases where singular configurations are included, the opposite is expected generically for a simple reason. We usually define
\begin{align}
Z=\int Dg A[g]
\end{align}
and the matter-gravity coupled integral $Z=\int Dg D\phi A[g,\phi]$ by specifying how to sum over the values of a gravitational variable $g$ on a lattice or some other structures. The case that $g$ does not assume any value somewhere is excluded in this step.

\section{Classical approximations}\label{sec:ca}

If singular spacetimes do not even arise in a generic gravitational path integral, how come that they are essential in studies of black hole and cosmology in classical gravity?

In a classical theory, what we care about is a differential equation -- the classical equation of motion. Singular spacetimes solve the differential equation for certain initial conditions, so are relevant. This is to be contrasted with quantum theory, where the path integral we care about is an integral. Instead of solving any differential equation, the relevant configurations are enumerated according to the values the gravitational variables take as explained above. This difference makes singular spacetimes irrelevant. 

On the other hand, the path integral receives dominating contributions from its stationary points, which are solutions to the Euler-Lagrange equation. Will singular spacetimes not become relevant in the leading-order approximation in light of this, even for the quantum theory?

The answer is no. What singular spacetimes solve are initial value problems. Yet the stationary points need to solve boundary value problems. Therefore singular spacetimes are still irrelevant for the leading approximation of the quantum theory.

\begin{figure}
    \centering
    \includegraphics[width=.65\textwidth]{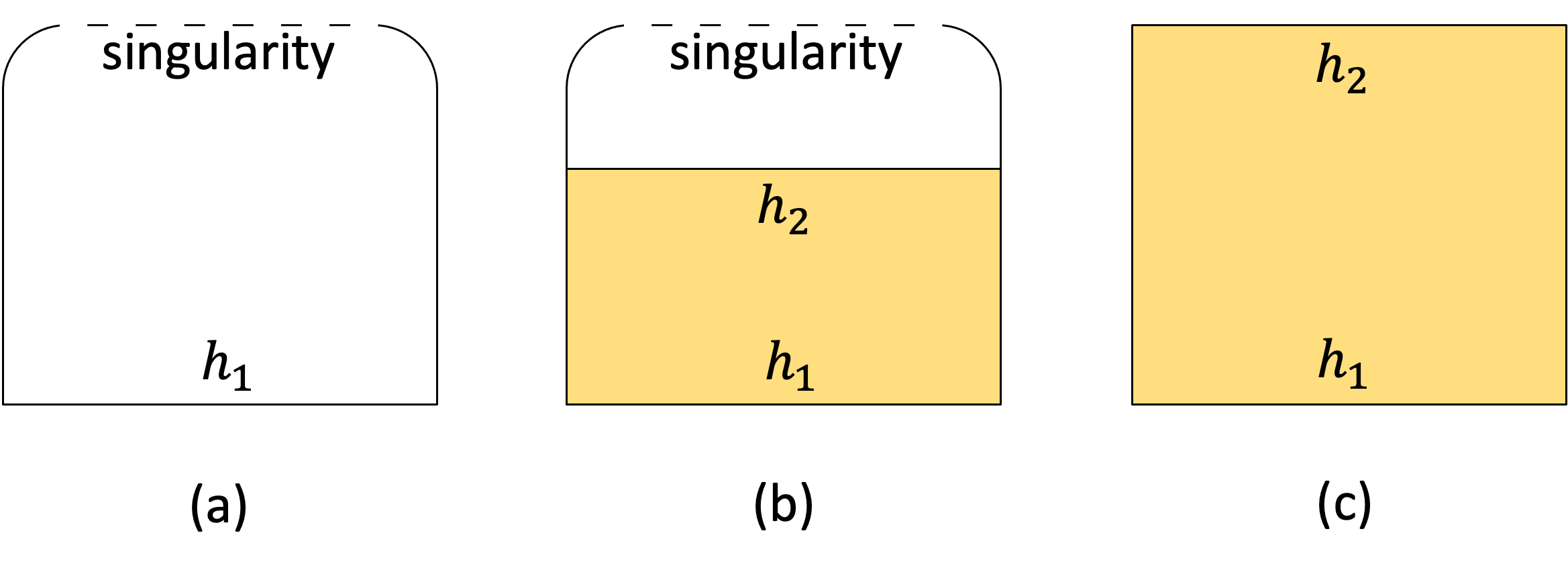}
    \caption{(a) A singular configuration with initial condition $h_1$ cannot match any final condition $h_2$. To also match some $h_2$, the configuration needs either to be truncated to remove the singularity, as in (b), or replaced by an entirely new configuration that shares $h_1$, as in (c).}
    \label{fig:pibcs}
\end{figure} 

To see this in more details, consider a path integral of the form
\begin{align}\label{eq:bpi}
Z[h]=\int_{h} Dg A[g],
\end{align}
where the sum obeys the condition $h$ on the boundary $B$ of the region. A special case of common interest is when $Z[h]$ gives the transition amplitude between $h_1$ and $h_2$, the two parts of $h$ on the two components $B_1$ and $B_2$ of $B$. In general, the stationary points $\overline{g}$ for \eqref{eq:bpi} need to solve the equation of motion for the boundary condition $h$, i.e., $\overline{g}|_B=h$. This condition cannot be met by singular spacetimes, because they cannot take the value $h$ on its total boundary. For instance, suppose $h$ decomposes into $h_1$ and $h_2$, where $h_1$ coincides with the initial boundary geometry for a portion of a Schwarzchild black hole. A singular spacetime may solve the equation of motion under the initial condition $h_1$ alone, but it can never match $h_2$ on the other part of the boundary (\Cref{fig:pibcs}).

We may be interested in path integrals of a more general form
\begin{align}\label{eq:bpi2}
Z[\psi]=\int D h \int_{h} Dg A[g] \psi[h].
\end{align}
A special case of interest is when $h$ decomposes into $h_1$ and $h_2$ as above, and we consider
\begin{align}\label{eq:bpi3}
Z[\psi]=\int D h_2 ~\phi[h_2]=\int D h_2 \int D h_1 \int_{h_1\cup h_2} Dg A[g] \psi[h_1],
\end{align}
where the complex function $\psi[h_1\cup h_2]=\psi[h_1]$ depends only on $h_1$. Here $\phi[h_2]$ is analogous to the wave function $\phi(x_2)=\int D x_1 \int_{x(t_1)=x_1}^{x(t_2)=x_2} Dx ~ A[x] \psi(x_1)$ of a non-relativistic particle at time $t_2$ given its wave function $\psi(x_1)$ at time $t_1$. Another special case of interest is when $\psi(h)=1$ identically which yields $Z$ as the full partition function. In general, the stationary points $\overline{g}$ for \eqref{eq:bpi2} are obtained by first solving $\delta_h (A[g]\psi[h])=0$ for the stationary points $h_S$ on the boundary, and then solving $\delta_g (A[g]\psi[h])=0$ for $\overline{g}$ under each boundary condition $h_S$. In cases such as $\psi[h_1\cup h_2]=\psi[h_1]$ when the boundary equation $\delta_h (A[g]\psi[h])=0$ is not very constraining, there could by many stationary points $h_S$. Yet $\overline{g}$ are obtained case by case, and in any case we solve $\delta_g (A[g]\psi[h])=0$ under a fixed boundary condition $h_S$. As explained in the last paragraph, singular spacetimes cannot be solutions.

Finally, coupling matter yields path integrals of the form \begin{align}
Z[\psi]=\int D h D\chi \int_{h,\chi} Dg D\phi A[g,\phi] \psi[h, \chi],
\end{align}
where $\chi$ are the boundary matter configurations. For the same reasons as in the pure gravity case, geodesically incomplete singular configurations do not solve the Euler-Lagrange equation to form stationary points.

\section{Tunneling and complex spacetimes}\label{sec:tcs}

Singular solutions are essential in classical gravity for understanding black holes and cosmology. If singular spacetimes become irrelevant in the quantum theory and in its classical approximation, what replaces them for understanding black holes and cosmology?

A short answer is that they are replaced by tunneling processes whose stationary points are complex-valued spacetimes that satisfy the boundary conditions \cite{Gielen2016QuantumSingularities, Chen2017FuzzySpace, Bramberger2017QuantumSingularities}. These spacetimes solve the classical equation of motion without being singular, as the singularity theorems are evaded by allowing the spacetimes to be complex-valued.

The situation is quite similar to particle tunneling. In the classical theory energy conservation prevents the particle trajectory from extending beyond a potential barrier, analogous to the singularity theorems preventing the spacetimes from extending beyond some finite affine parameters along causal paths \cite{wald2010general}. However, in the quantum path integral the boundary conditions for detecting the particle beyond the potential barrier yield positive probabilities, accompanied by complex-valued solutions to the classical equation of motion \cite{Turok2014OnTime, ChermanReal-TimeInstantons, Tanizaki2014Real-timeTunneling}. The ban of energy conservation is evaded, because complex trajectories come with complex momentum so that the kinetic energy can be negative under the potential barrier.

While the basic scheme of this tunneling picture for singularity resolution is clear, some open problems remain to be clarified. Firstly, how should we understand the complex spacetimes that arise in the classical approximations to singularity-resolving processes, when the original path integral includes only real-valued spacetimes in the sum? Mathematically, the complex stationary points can be understood through Picard-Lefschetz theory as belonging to the deformed integration contours in the complexified domain \cite{Feldbrugge2017LorentzianCosmology}. Yet physically, an intuitive understanding of the complex spacetimes seems to be missing. In particular, does a complex spacetime come with a causal structure? Are there lightcones on it? How do matter and information propagate in it? Secondly, the above qualitative picture holds for generic boundary conditions. Yet which specific boundary conditions should we use for the black holes and the cosmos of our world? Is this purely an empirical question, or are there principles that guide the choices? In a theory where the trivial idea of singularity resolution applies, these constitute some interesting questions to be investigated further.

\section{Comments on some alternative views}


I have presented a view that singularity resolution is trivial in gravitational path integrals because singular spacetime configurations are generically absent in the sum. Given an explicitly defined gravitational path integral one could check if this view applies. When it does, the interesting tasks are to understand gravitational tunneling processes and their boundary conditions, as well as the physical meaning of the corresponding stationary points which are complex-valued.

From this perspective, current attempts at using \textit{real-valued} regular spacetimes to give effective descriptions of singularity-resolving processes along the lines of \cite{Hayward2006FormationHoles} could be misguided. Should an individual spacetime be chosen to capture the essence of the quantum processes, a complex-valued spacetime is more appropriate than a real-valued one. The phenomenological consequences of complex-valued spacetimes for gravitational waves and black hole images are certainly also worth studying.

The trivial idea of singularity resolution is also worth discussing in view of some contrary statements present in the literature. For example, in a relatively recent work \cite{Borissova2020TowardsIntegral} it is stated that in a gravitational path integral based on the metric variable $g_{ab}$,
\begin{quote}
all possible metric configurations (modulo diffeomorphisms) are being summed over. Thus, the singular spacetime metrics that constitute solutions of the field equations in GR are included in the path integral.
\end{quote}
The problem with metric-variable gravitational path integral is that it is unclear how to define it in a non-perturbative and Lorentzian setting. In quantum field theories, the standard way to specify a path integral non-perturbatively is through lattices \cite{Peskin1995}. This route leads back to theories such as those discussed in \Cref{sec:scdnb}. For the theories considered there singular configurations are seen to not belong to the path integral. Alternatively one might consider functional renormalization group for an in-principle non-perturbative specification of an Euclidean gravitational path integral. However it is unclear how to take this approach over to the Lorentzian setting \cite{Bonanno2020CriticalGravityb} which is suitable for discussing spacetime singularities. Without being referred to a Lorentzian path integral with an explicitly defined measure, it is difficult to be convinced of the claim that singular spacetime configurations are included in the path integral. 

As another example from recent works, in \cite{ChenSolvingIntegral} it is stated as an ``essential condition'' for a solution to the ``information loss paradox'' that for an Euclidean gravitational path integral (EPI):
\begin{quote}
There exist at least two histories, say $h_1$ and $h_2$, that contribute to EPI, where $h_1$ is an information-losing history while $h_2$ is an information-preserving history.
\end{quote}
Here ``information-losing history'' means ``the semi-classical history of an evaporating black hole in which the unitary evolution would be lost when the black hole has completely evaporated'' \cite{ChenSolvingIntegral}, and from Figure 1 of that paper one might infer that an ``information-losing history'' is geodesically incomplete. The main idea of \cite{ChenSolvingIntegral} is to understand information propagation of quantum black holes as tunneling processes in gravitational path integrals. It seems the main points of \cite{ChenSolvingIntegral} depends on the presence of an information-preserving history $h_2$ rather than an information-losing history $h_1$, and for reasons discussed in \Cref{sec:tcs} I remain hopeful that the program of \cite{ChenSolvingIntegral} to understand black hole information topics through gravitational tunneling will succeed. 

The present trivial idea also differs from non-trivial ideas such as those listed in \Cref{sec:i}. What accounts for the differences? Firstly, the trivial idea is based on gravitational path integrals, whereas ideas such as \cite{Ashtekar2006QuantumInvestigation, Ashtekar2009SingularityOverview, PerryFutureParadox} are based on Wheeler-DeWitt type models or theories. Secondly, the trivial idea refers to explicitly defined path integrals, whereas ideas such as \cite{Borissova2020TowardsIntegral, Horowitz2003TheState} refer only to formal expressions of path integrals.  Without an explicitly specified measure it is impossible to check if singular configurations belong to the path integral sum. In view of the present work, this could be an important missing step that changes the conclusion on how singularities are resolved in quantum gravity.

\section*{Acknowledgement}

I thank Johanna Borissova and Qiaoyin Pan for an interesting discussion that triggered this note, and Johanna Borissova and Achim Kempf for comments on an earlier draft. I am very grateful to Lucien Hardy and Achim Kempf for long-term encouragement and support. Research at Perimeter Institute is supported in part by the Government of Canada through the Department of Innovation, Science and Economic Development Canada and by the Province of Ontario through the Ministry of Economic Development, Job Creation and Trade. 

\bibliographystyle{unsrt}
\bibliography{mendeley.bib}

\end{document}